\documentclass[%
aip,
amsmath,amssymb,
reprint,%
]{revtex4-1}

\usepackage{graphicx}% Include figure files
\usepackage{dcolumn}% Align table columns on decimal point
\usepackage{bm}% bold math
\usepackage[utf8]{inputenc}
\usepackage[T1]{fontenc}
\usepackage{mathptmx}
\usepackage{etoolbox}

\usepackage[english]{babel}

\usepackage{siunitx}
\usepackage{import}
\usepackage[hidelinks]{hyperref}
\usepackage[capitalise]{cleveref} %,noabbrev

\makeatletter
\def\@email#1#2{%
	\endgroup
	\patchcmd{\titleblock@produce}
	{\frontmatter@RRAPformat}
	{\frontmatter@RRAPformat{\produce@RRAP{*#1\href{mailto:#2}{#2}}}\frontmatter@RRAPformat}
	{}{}
}%
\makeatother
\begin{document}
	
	\preprint{AIP/123-QED}
	
	%\title[Mechanical design of optical cavity support structure for ultra-high vacuum application]{Mechanical design of optical cavity support structure for ultra-high vacuum applications}
	% Force line breaks with \\
	\title{Vibration damping platform for cavity quantum-electrodynamics experiments}
	\author{N. Sauerwein}
	\affiliation{Institute of Physics, \'Ecole Polytechnique F\'ed\'erale de Lausanne (EPFL), CH-1015 Lausanne, Switzerland}%Lines break automatically or can be forced with \\
	\email{nick.sauerwein@hotmail.de}
	
	\author{T. Cantat-Moltrecht}%
	
	\affiliation{Institute of Physics, \'Ecole Polytechnique F\'ed\'erale de Lausanne (EPFL), CH-1015 Lausanne, Switzerland}%

	\author{I. T. Grigoras}%
	
	\affiliation{Institute of Physics, \'Ecole Polytechnique F\'ed\'erale de Lausanne (EPFL), CH-1015 Lausanne, Switzerland}%
	
	\author{J.-P. Brantut}
	\homepage{https://lqg.epfl.ch}
	\affiliation{Institute of Physics, \'Ecole Polytechnique F\'ed\'erale de Lausanne (EPFL), CH-1015 Lausanne, Switzerland}%
	
	\date{\today}% It is always \today, today,
	
	\begin{abstract}
		We present a mechanical platform with enhanced vibration damping properties for cavity quantum-electrodynamics experiments. It is based on a composite design that combines a soft, vibration-damping core with a rigid shell maintaining optical alignment. It passively damps the vibrations generated by piezoelectric actuators controlling the mirror positions. The mechanical resonances of the platform, which lead to a length change of the cavity are efficiently suppressed up to \SI{100}{\kHz}. Our platform is ultra-high vacuum compatible and can be used in most applications, in particular where long cavities and optical access to the cavity center are required.
	\end{abstract}
	
	\maketitle
	
	Every mechanical structure is subject to vibrations, either originating from its environment, or excited within the structure by the inertia of moving parts. High-finesse optical interferometers are among the devices most sensitive to vibrations: displacements of the mirrors at the picometer scale lead to significant changes in the cavity transmission. This sensitivity can be exploited for sensing \cite{aspelmeyerCavityOptomechanics2014, ligoscientificcollaborationandvirgocollaborationObservationGravitationalWaves2016}, but for most applications this leads to significant disturbances, in particular for cavity quantum-electrodynamics (cQED) experiments which require absolute stability of the cavity length. Even though shielding against external disturbances can be efficiently achieved through the use of well established techniques \cite{okanoVibrationIsolationScanning1987}, the need for active stabilization of the cavity length implies the use of actuators on the mirrors for continuous feedback, triggering vibrations in the structure. In particular, sharp vibration resonances severely limit the bandwidth of feedback loops and ultimately represent the main limitation to cavity stability. This is especially relevant for new generations of cQED experiments employing long cavities or complex geometries \cite{slamaSuperradiantRayleighScattering2007,bertoldiSituCharacterizationOptical2010,kesslerOptomechanicalAtomcavityInteraction2014,kollarAdjustablelengthCavityBose2015a,leonardSupersolidFormationQuantum2017,nguyenOperatingNearconcentricCavity2018,davisPhotonMediatedSpinExchangeDynamics2019,rouxCavityassistedPreparationDetection2021,jaffeAberratedOpticalCavities2021}
, where mechanical resonances can occur at low frequencies.
	
	Damping and mounting strategies have been implemented for cavity systems, for example in the context of laser stabilization or interferometers \cite{brilesSimplePiezoelectricactuatedMirror2010, notcuttSimpleCompact1Hz2005,chadiNoteSimpleCompact2013,sinclairInvitedArticleCompact2015,andersonComparisonVibrationDamping2008}. However, the constraints of an ultra-high vacuum (UHV) environment have prevented their direct use in cQED applications. Advanced signal filtering methods have therefore been developed to cancel mechanical resonances within feedback loops \cite{ryouActiveCancellationAcoustical2017}. In this note, we present the design of a composite cavity holder that features efficient passive damping while maintaining UHV compatibility. Compared to a bulk platform with the same geometry, our composite platform shows efficient suppression of all mechanical vibration modes up to \SI{100}{\kHz} and a smooth phase response compatible with high-bandwidth feedback loops.
	
	\begin{figure}[ht!]
		\centering
		\includegraphics[width = \linewidth]{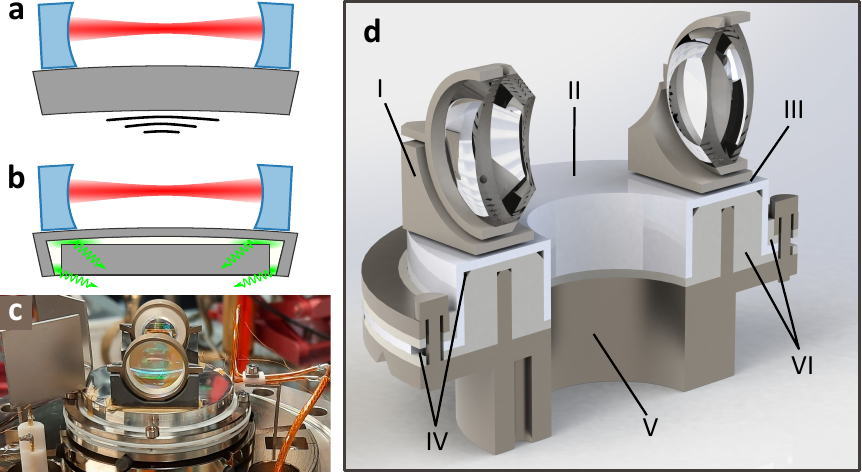}
		\caption{\textbf{Principle and design of the cavity holder.} Schematic \textbf{a} shows the fundamental mode of vibration of a bulk cavity holder which leads to a shift of the optical resonance frequency. The holder in \textbf{b} is damped (dissipation in green), suppressing mechanical vibrations. Picture \textbf{c} of the final assembly of the optical cavity was taken before the system was installed inside the vacuum chamber. Rendering \textbf{d} represents a section of the cavity holder. The roman number annotations are referenced in the main text.}
		\label{fig:design}
	\end{figure}
	The cavity holder consists in a toroidal platform, hosting two shear piezoelectric actuators (see \cref{fig:design}d III) onto which the cavity mirrors (\cref{fig:design}d I) are glued. Its geometry is chosen to maximize the optical access from all directions. The motion of the mirrors caused by the cavity length stabilization loop drives mechanical vibration modes of the holder, as illustrated in \cref{fig:design}a. With a cavity mirror spacing of roughly \SI{2.5}{\cm}, the lowest vibration mode frequency lies in the \SI{10}{\kHz} range, weakly dependent on the choice of materials among vacuum compatible and machinable metals. 
	
	In order to damp these resonances, we designed the holder as a composite core-shell structure comprising three different pieces: (i) a rigid u-shaped shell maintaining the alignment of the cavity mirrors (\cref{fig:design}d II), (ii) a soft, thick damper filling most of the shell's volume (\cref{fig:design}d VI) and (iii) a high-rigidity base pressed inside the damper (\cref{fig:design}d V). The full 3D model of the cavity assembly is publicly accessible on zenodo \cite{sauerwein3DModelVibration2021}. The overall operation principle is illustrated in \cref{fig:design}b: the vibration mode driven by the piezoelectric actuators deforms the shell, which transmits the deformation to the damper. The damper is thus compressed against the base, efficiently dissipating energy within its volume. 

	\paragraph*{Materials}

\begin{figure}[h]
		\centering
		\includegraphics[width = \linewidth]{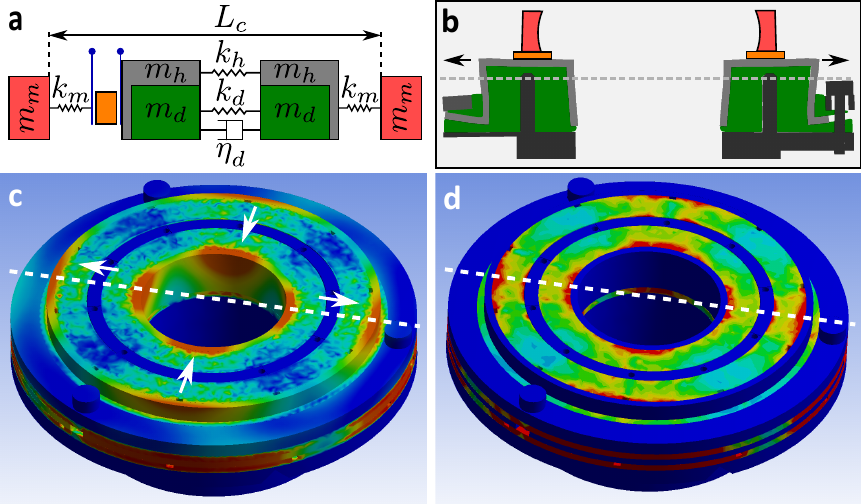}
		\caption{\textbf{Model and simulation results.} Drawing \textbf{a} depicts a simple spring model of the cavity assembly. Drawing \textbf{b} shows the exaggerated displacement of the cavity holder under the main mechanical resonance (colors as in \textbf{a}). In \textbf{c} and \textbf{d} respectively, the total displacement and the von Mises strain in the holder assembly are shown, in a section taken along the gray dashed line in \textbf{b}. The white dashed lines indicate the cavity axis and the white arrows represent the direction of deformation.}
		\label{fig:simple_model}
	\end{figure}

	The simple mass-spring model presented in \cref{fig:simple_model}a and b helped us extract the main figures of merit for choosing the materials. In this model, the mirrors (red) are connected to the holder (grey) through springs of strength $k_m$. The mirror-holder distance is tuned by a piezoelectric actuator (orange). The holder is fixed to the damper (green), which has a damping coefficient $\eta_d$. A simple calculation shows that in the limit of stiff and light mirrors, the resonance frequency of the ensemble is proportional to $ \sqrt{\widetilde{k_h}/\widetilde{m}} $, the square-root of the effective specific stiffness, where $\widetilde{k_h} = k_h + k_d$ is the combined elastic modulus of the holder and damper, and $\widetilde{m} = m_m + m_h + m_d$ is the combined mass of the mirror, holder and damper. The quality factor $Q$ of this resonance is proportional to $\sqrt{\eta_d \widetilde{m}}$.
	
	Minimizing the effect of mechanical vibrations requires high resonance frequencies, to allow for higher stabilization bandwidth, and low $Q$-factors, implying the use of materials with high specific stiffness and high damping. Unfortunately, all of the common stiff materials intrinsically have low damping coefficients, hence the use of a composite structure.
	
	For the top shell, we chose aluminum as a compromise between machinability and high specific stiffness \footnote{Taking simple precautions, its UHV compatibility can be ensured \cite{hoffmanHandbookVacuumScience1997}}. For the damper, we chose Teflon (PTFE) for its high damping coefficient at room temperature, due to two first-order crystal-crystal transitions at \SI{19}{\degreeCelsius} and \SI{30}{\degreeCelsius} \cite{lehnertComparativeQuantitativeStudy1997, raePropertiesPolyTetrafluoroethylene2004}. It can be machined easily and is UHV compatible, as opposed to most rubber materials commonly employed as dampers. Additionally, it can be baked at temperatures exceeding \SI{150}{\degreeCelsius}. For the base, we chose titanium for its high stiffness,
	machinability and UHV compatibility.
	
	\paragraph*{Geometry}
	
	We optimized the geometry of the holder using finite element (FE) simulations \cite{AnsysAcademicResearch}. With a modal analysis, the shape of the top plate was optimized for high resonance frequencies (\textgreater\SI{20}{\kHz}) and low weight. The mode that changes the cavity length the most is the fundamental stretch mode (see \cref{fig:simple_model}b,c and d). A cylindrical u-shaped top plate was found to maximize its frequency, making it similar to that of a bulk holder, while reducing the weight significantly.
	
	Because damping is proportional to the time-variation of strain, a key parameter here is the efficient coupling of the motion of the top plate to the compression of the damper. To achieve maximal strain in the damper for a given deformation of the holder, we use a rigid base pressed inside the bulk of the damper. This way, the deformation of the top plate compresses the damper against the base, as illustrated in \cref{fig:simple_model}b. \cref{fig:simple_model}c represents the distribution of the absolute value of the total deformation of our cavity holder for the stretch mode, obtained using a FE simulation. The largest deformations appear on the outer shell and leave the base (blue circle in the middle) fixed, illustrating the compression mechanism of the damper between the shell and the base. The corresponding strain distribution is presented in \cref{fig:simple_model}d, confirming the efficient strain transfer to the damper.
	
	\paragraph*{Assembly}
	
	The top plate consists in a \SI{1}{\mm} thick aluminum shell, encapsulating the PTFE damper, which is pressed against the titanium base. The parts are fastened together using titanium screws at the outer edges (see \cref{fig:design}d)\footnote{There is no direct, rigid contact between the top plate and the base thanks to a damping PTFE ring on the outside}. Apart from the piezoelectric actuators for cavity length stabilization, there are no moving parts, minimizing the risk of spurious vibrations. The assembly itself rests on a stack of two non-magnetic stainless steel rings separated by Viton rods for isolation \cite{okanoVibrationIsolationScanning1987}, themselves sitting on the bottom flange of a vacuum chamber, as shown in \cref{fig:design}c. All the materials are inexpensive, widely available and  were machined easily using standard mechanical tools. A network of venting channels in the damper and base (see \cref{fig:design}d IV) prevents virtual leaks for UHV operation.
	
	\paragraph*{Mechanical response measurements}
	\label{sec:measurements}

	We measured the frequency-dependent mechanical response of the holder by exciting harmonic vibrations and measuring them with a Michelson interferometer, using a mirror glued to the holder. With a slow feedback loop (\SI{200}{\Hz} bandwidth), the interferometer was stabilized to a half-fringe configuration. Harmonically driving the piezoelectric actuator and demodulating the interference signal allowed to retrieve the complex transfer function relating the mechanical displacement of the mirror to the actuator voltage. To evaluate the performance of our damped cavity holder, we compared its response with that of a geometrically identical holder made of bulk stainless steel.
	
	\begin{figure}[ht]
		\centering
		\includegraphics[width = \linewidth]{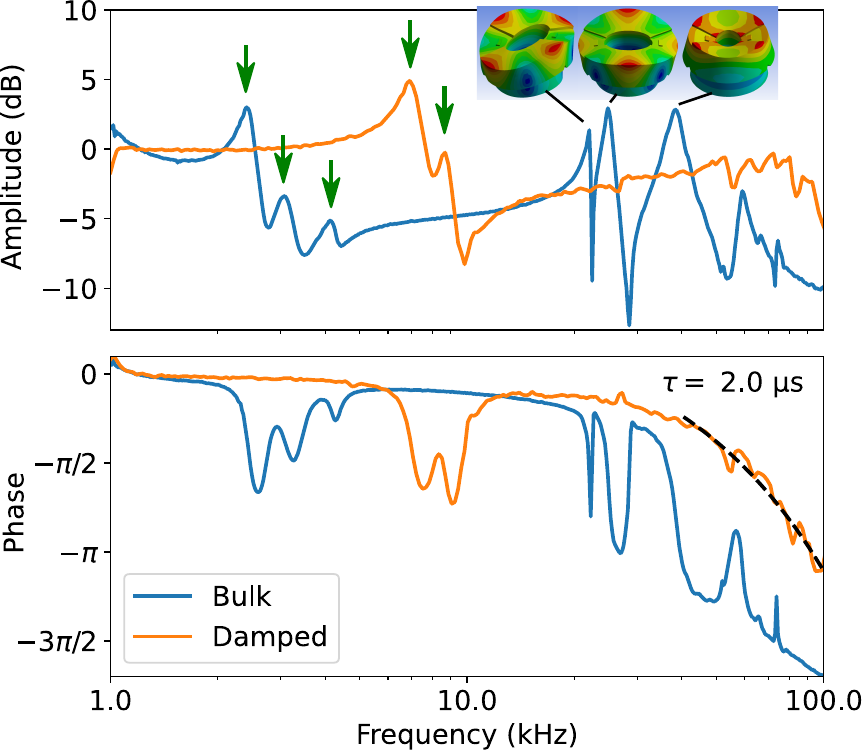}
		\caption{\textbf{Mechanical response functions.} The graphs present the amplitude and phase of mechanical response functions for two different holder designs (a damped holder and a geometrically identical bulk stainless steel version), measured with a Michelson interferometer The inset shows the exaggerated deformations of the bulk holder for the the indicated resonances.}
		\label{fig:measurements}
	\end{figure}
	
	The results are presented in \cref{fig:measurements}. For both holders, a set of isolated resonances at frequencies lower than \SI{10}{\kilo \Hz}, indicated by arrows in \cref{fig:measurements}, results from the flexure modes of the mirror itself. We have confirmed this first by an ab-initio FE calculation which we expect to be quantitatively accurate for such a simple homogeneous part and second by changing the weight of the mirror, which led to a change of the mirror resonance frequencies. 

	A second set of resonances at higher frequencies, insensitive to changes of the mirror mass, results from the modes of the holder itself. These resonances are very prominent in the bulk stainless-steel holder, appearing as sharp peaks and abrupt phase jumps by about $-\pi$. A comparison with the mode frequencies calculated from the FE model allows us to identify them as flexure and stretch modes of increasing order. 
	
	In contrast, the composite holder is essentially resonance-free up to \SI{100}{\kHz}. In particular, the phase of its mechanical response stays above $-\pi/2$ up to \SI{60}{\kHz}, and smoothly rolls with a typical timescale of \SI{2}{\micro \second}, making the platform suited for high-bandwidth regulation of the cavity length. At these higher frequencies, vibration modes become dense and overlapping, making electronic filtering challenging \cite{ryouActiveCancellationAcoustical2017}, while the damping properties of our composite holder do not degrade.
	
	\paragraph*{Vacuum compatibility}
	
	The holder has been integrated in a cQED setup, with a pair of mirrors forming a high-finesse cavity. After thorough cleaning and pre-baking of the different parts of the holder at their maximum allowed temperature and under vacuum, the holder was assembled with cavity mirrors under ambient conditions, and installed in a standard cold atoms UHV system. In our experiment, we use a nearly-concentric cavity, which is sensitive to misalignment at the micrometer scale. Therefore, to prevent thermal deformation, the aligned cavity system was not baked after installation. Nevertheless, we measure a pressure in the low \SI{e-11}{\milli \bar} range, and a lifetime of laser cooled atoms of \SI{32}{\second} at the center of cavity which increased to \SI{68}{\second} after less than a year of operation, demonstrating the full UHV compatibility of the holder and its suitability for the most demanding quantum gas experiments. 
	
	We acknowledge the technical assistance of Gilles Grandjean, Luc Chevalley and Adrien Grisendi. This work is supported by the European Research Council (ERC) under the European Union’s Horizon 2020 research and innovation programme (grant agreement No 714309), the Swiss National Science Foundation (grant No 184654), the Sandoz Family Foundation-Monique de Meuron program for Academic Promotion and EPFL.
	
	\bibliographystyle{naturemag}
	\bibliography{Cavity_holder_paper}
	
	%\nolinenumbers
	
\end{document}